\documentclass[letterpaper, 10 pt, conference]{IEEEtran}  

\IEEEoverridecommandlockouts                              
\usepackage{hyperref}
\usepackage{physics}
\usepackage{svg}
\usepackage[clean=false]{svg-extract}
\usepackage{tikz}
\usepackage{tikzscale}
\usepackage{pgfplots}
\usetikzlibrary{matrix}
\usepgfplotslibrary{groupplots}
\pgfplotsset{compat=newest}
\usepgfplotslibrary{external}
\newcommand{\minimize}[1]{\underset{#1}{\text{minimize}}}
\newcommand{\st}{\text{subject to}}
\newcommand{\isopsi}{\tilde \psi}
\newcommand{\isoU}{\widetilde{U}}

\newcommand{\iso}{\text{iso}}

\usepackage{amsmath} 
\usepackage{amssymb}  
\usepackage{graphicx}
\usepackage[square,numbers,compress]{natbib}
\title{\LARGE \bf
    Direct Collocation for Quantum Optimal Control
}

\author{Aaron Trowbridge,$^1$ Aditya Bhardwaj,$^2$ Kevin He,$^2$ David I. Schuster,$^{2,3}$ and Zachary Manchester$^1$
\thanks{$^1$Carnegie Mellon University, Robotics Institute}
\thanks{$^2$University of Chicago, Department of Physics}
\thanks{$^3$Stanford University, Department of Applied Physics}
}

\begin{document}

\maketitle
\thispagestyle{empty}
\pagestyle{empty}

\begin{abstract}
  We present an adaptation of direct collocation – a trajectory optimization method commonly used in robotics and aerospace applications – to quantum optimal control (QOC); we refer to this method as Pade Integrator COllocation (PICO). This approach supports general nonlinear constraints on the states and controls, takes advantage of state-of-the-art large-scale nonlinear programming solvers, and has superior convergence properties compared to standard approaches like GRAPE and CRAB.  PICO also allows for the formulation of novel free-time and minimum-time control problems – crucial for realizing high-performance quantum computers when the optimal pulse duration is not known \textit{a priori}. We demonstrate PICO’s performance both in simulation and on hardware with a 3D circuit cavity quantum electrodynamics system.
  
  \hfill

\end{abstract}



\begin{figure}[h]
\includegraphics[width=0.4832\textwidth]{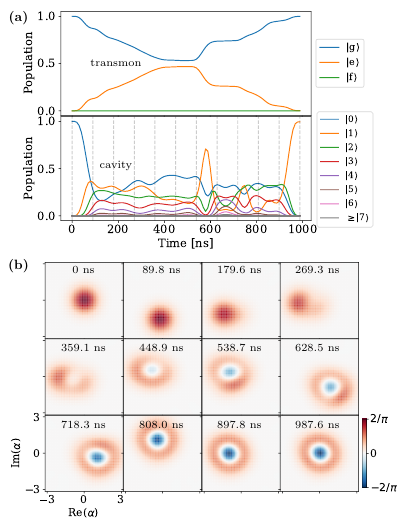}
\caption{
\textbf{PICO pulse simulated state evolution for a target state of $\ket{g1}$}. \textbf{(a)} Time evolution of the separate transmon and cavity levels for the duration of the control pulse. 
Cavity photon number populations increase in order, before being manipulated to end in the single photon state. For this pulse, the transmon $\ket{f}$ state and cavity photon numbers above $\ket{6}$ are not significantly populated. Dashed vertical lines indicate the times of the Wigner tomography slices in (b).
\textbf{(b)} Time evolution of the cavity populations in phase space, shown as Wigner tomography plots. Starting in the vacuum $\ket{0}$ photon state, the control pulse evolves the system to the final $\ket{1}$ photon state. Time slices go from left-right and up-down at the equally spaced time points indicated by the gray vertical lines in the cavity populations plot in (a).
} 
\label{hardware_figure_p1}
\end{figure}

\section{Introduction}
The field of optimal control, which has its origins in aerospace engineering and robotics, has produced a large body of sophisticated methods for solving control problems fundamentally similar to the problems posed in quantum optimal control (QOC) \cite{manchesterTrajOpt, altro, howell2023direct, jackson2021attitude, underactuated}. However, many of these methods have yet to be adopted by those working on control problems for quantum systems. This work aims to bridge the gap between robotic control and quantum control and, in so doing, provide a new perspective to practitioners of QOC. 

Current widely used QOC methods \cite{khaneja2005grape, caneva2011crab} fall into the category of \textit{indirect methods}, in that they do not include the states as decision variables.
In this work, we introduce Pad\'e Integrator COllocation (PICO) as an alternative, fully direct method. PICO is an adaptation of the direct collocation method \cite{hargraves1987dircol} tailored to the quantum setting. With this method, we are able to leverage large-scale nonlinear programming solvers, specifically the interior point method IPOPT \cite{wachter2006ipopt}.  We demonstrate that PICO, due to being a direct method, improves upon existing indirect methods in several ways, and is able to produce state-of-the-art results in both simulation and on hardware.  

Our contributions include:
\begin{itemize}
    \item A novel formulation of QOC using direct collocation
    \item Structure-preserving \textit{Pad\'e integrators} that efficiently compute an approximation of the matrix exponential to simulate quantum dynamics
    \item An open-source implementation QuantumCollocation.jl \cite{Trowbridge_QuantumCollocation_jl_2023}
\end{itemize}
We demonstrate state-of-the-art results, both with regard to fidelity and time optimality, on a set of simulated problems of increasing difficulty. We also show that PICO produces pulses with state-of-the-art performance on an experimental system.

This paper is structured as follows: First, in Sec.~\ref{sec: back}, we review the problem of quantum optimal control and existing solution methods.  Then, in Sec.~\ref{sec: qcp}, we introduce our method, PICO. In Sec.~\ref{sec: example} we demonstrate PICO's performance on three QOC problems of increasing difficulty. Finally, in Sec.~\ref{sec: hwr}, we present the hardware results achieved with PICO. 
\newpage

\section{Background} \label{sec: back}

In this section, we review relevant prior work.  First, we discuss the mathematical formulation of quantum optimal control. Then, we review gradient-based methods for solving QOC problems, followed by gradient-free methods. Finally, we review the direct collocation method commonly employed to solve trajectory optimization problems in aerospace and robotics applications \cite{Hargraves87a,Betts92,Pardo15a,Posa16a}.

\subsection{Quantum Optimal Control}
The control of quantum systems can be framed as an optimization problem over the space of time-dependent state trajectories subject to dynamics described by the Schr\"odinger equation. For a unitary operator, in our setting, this is given by:
\begin{equation} \label{eq:1}
   \dot U = -i H(\vb{a}(t)) U.
\end{equation}
\noindent

The dynamics are controllable via \textit{drive} parameters $\vb{a}(t) \in \mathbf{R}^d$ in the Hamiltonian.  For simplicity, we will limit ourselves to considering Hamiltonians of the form,
\begin{equation}
  H(\vb{a}(t)) := H_0 + \sum_i a_i(t) H_i,
\end{equation}
\noindent
where $t \in [0, T]$, $H_0$ is the system's \textit{drift} term, and $H_i$ are referred to as the \textit{drive} terms.  Handling other types of Hamiltonians---e.g. those that are nonlinear in the controls---is also possible in our method.

QOC problems typically fall into three categories, corresponding to three different types of quantum objects: quantum states $\ket{\psi(t)}$, which satisfy $\partial_t\ket{\psi} = -iH\ket{\psi}$; unitary operators $U(t)$, which satisfy Eqn.~(\ref{eq:1}); and density operators $\rho(t)$, which satisfy $\dot\rho(t) = -i\qty[H, \rho]$.  

QOC methods are agnostic to the form of the state and the dynamics, so we will primarily discuss unitary operators, as they are arguably the most general. In this case, given a desired gate $U_{\text{goal}}$, the objective we will be most concerned with is the unitary infidelity loss, defined as,
\begin{equation} \label{eq: trace_fid}
  \ell(U) := 1 - \frac{1}{n}\abs{\text{tr}\qty(U_{\text{goal}}^\dagger U)},
\end{equation}
for $U \in SU(n)$.

We discretize the time interval $[0, T]$ into $N$ time steps of size $\Delta t$; the states and controls at each time step are denoted $U_k$ and $\vb{a}_k$, respectively. One then solves the following optimization problem: 
\begin{equation}
\begin{split}
  \minimize{\vb{a}_{1:N-1}} \quad &J(\vb{a}_{1:N-1}) = \ell(U_N(\vb{a}_{1:N-1})),
\end{split}
\end{equation}
\noindent
where 
\begin{equation} \label{eq: unitary rollout}
  U_N(\vb{a}_{1:N-1}) = \prod_{k=1}^{N-1} \exp\qty(-i H(\vb{a}_k) \Delta t).
\end{equation}
\noindent

Current approaches for solving this problem fall into two categories: \textit{gradient-based} methods and \textit{basis function} methods.  Both of these are \textit{indirect} methods, as they treat the final state $U(T)$ as a function of the controls and minimize the objective $J(\vb{a}_{1:N-1})$ only over the controls $\vb{a}_{1:T-1}$, as opposed to considering both the states and controls as decision variables in what are known as \textit{direct} methods \cite{Betts98a}. 

\subsection{Gradient-Based Methods}

Gradient-based methods involve initializing the controls $\vb{a}_{1:T-1}$ with an initial guess, rolling out the state using Eqn.~(\ref{eq: unitary rollout}), and then iteratively updating the controls using a gradient descent algorithm; this type of approach is also referred to as a \textit{shooting method}. Evaluating the objective $J(\vb{a}_{1:N-1})$, which requires a costly full rollout, is necessary at each iteration to compute a gradient and update the controls:
\begin{equation}
  \vb{a} \leftarrow \vb{a} - \beta \nabla J(\vb{a}).
\end{equation}

There are efficient ways to compute this gradient \cite{leung2017gpugrape}, but the approach is still limited by rollouts and other factors: The solution is dependent on the initial guess, gradient-based methods are prone to falling into local minima, and it is difficult to rigorously enforce constraints on the states.

The most popular gradient-based method is known as GRAPE (GRadient Ascent Pulse Engineering) \cite{khaneja2005grape}, which is available through the popular QuTiP python library. The company Q-CTRL also implements its own gradient-based optimization tool \cite{boulder_opal1, boulder_opal2}, which is the industry standard, and is what we compare our results to in this paper.

\subsection{Gradient-Free Methods}

There are other approaches that do not require taking gradients (but still require objective evaluations). These approaches typically utilize a parameterized basis of functions to sufficiently reduce the number of decision variables so that gradient-free optimization algorithms --- e.g. the Nelder-Mead simplex method --- can be utilized. In the case of the popular CRAB algorithm \cite{caneva2011crab}, this is accomplished by utilizing a Fourier basis for each pulse component:
\begin{equation}
  a(t) = 1 + \frac{\sum_{n=1}^{N_c} b_n \sin(\omega_n t) + c_n \cos(\omega_n t)}{\lambda(t)}.
\end{equation}
\noindent
Where $N_c$ is the number of terms taken in the series expansion and $\lambda(t)$ is chosen to enforce the boundary conditions. The result is a smaller optimization problem:
\begin{equation}
  \minimize{b_{1:N_c}, c_{1:N_c}, \omega_{1:N_c}} \quad J(b_{1:N_c}, c_{1:N_c}, \omega_{1:N_c}).
\end{equation}

\subsection{Trajectory Optimization and Direct Collocation}


An alternative approach for solving trajectory optimization problems is \textit{direct collocation} (DIRCOL) \cite{hargraves1987dircol}. DIRCOL is a gradient-based direct method that overcomes many of the limitations of indirect methods by including both the states and controls at each time step as decision variables, denoted by $x_k$ and $u_k$ respectively.  In this formulation, objective evaluations are cheap, as they do not involve rollouts, and the dynamics are explicitly enforced as equality constraints between \textit{knot points} (state and control samples) $z_k = (x_k, u_k)^\top$.  

Enforcing the dynamics as constraints is a key property of DIRCOL that allows numerical solvers to temporarily violate these constraints during intermediate steps of the solution process \textit{en route} to satisfying them at convergence.  This \textit{infeasible-start} capability, along with the ability to easily enforce constraints on the state variables, is the source of many of the advantages of direct over indirect methods. 

A trajectory optimization problem can be simply stated in the direct framework: We begin in an initial state, i.e. $x_1 = x_{\text{init}}$, and find a control sequence $u_{1:N-1}$ such that $x_N$ minimizes an objective $J$ consisting of a loss $\ell(x_N)$ measuring the distance between the final state $x_N$ and the goal state $x_{\text{goal}}$. We may also include other objectives terms, including e.g. penalties on higher derivatives of the control pulse to encourage smoothness.  It is common to enforce the dynamics constraints implicitly, i.e. as $f(z_k, z_{k+1}) = 0$. We can then write the DIRCOL problem as:
\begin{equation}
\begin{split}
  \minimize{z_{1:N}} \quad & J(z_{1:N})\\
  \st \quad & f(z_k, z_{k+1}) = 0, \\
            & x_1 = x_{\text{init}} ,
\end{split}
\end{equation}
which is a large sparse nonlinear program that can be efficiently solved with a nonlinear solver such as IPOPT.

\section{Pad\'e Integrator Collocation} \label{sec: qcp}

This section deals with formulating QOC problems as direct collocation trajectory optimization problems. We will focus on optimizing for $SU(n)$ gates. Using the loss $\ell(\cdot)$ defined in Eqn.~(\ref{eq: trace_fid}) and the naive dynamics
\begin{equation} \label{eq: quantum-dynamics-naive}
f(U_{k+1}, U_k, \vb{a}_k, \Delta t) = U_{k+1} - \exp(-i H(\vb{a}_k) \Delta t) \ U_k,
\end{equation} 
\noindent
where $\Delta t$ is fixed. A simple DIRCOL formulation of this problem can be written as follows:
\begin{equation} \label{prob: qoc-base}
\begin{split}
  \minimize{U_{1:N}, \vb{a}_{1:N}} \quad & \ell(U_N) \\
  \st \quad & f(U_{k+1}, U_k, \vb{a}_k, \Delta t) = 0 \\
            & U_1 = I_n 
\end{split}
\end{equation}

We detail several practical considerations:  First, we will cover how to convert the complex-valued objects in Eqn.~(\ref{prob: qoc-base}) to real values.  Next, we will discuss a novel, efficient way to enforce the dynamics, which avoids costly evaluations of the matrix exponential present in (\ref{eq: quantum-dynamics-naive}). Finally, we will discuss a few extensions for achieving \textit{smooth} and \textit{time-optimal} solutions.

\subsection{Isomorphic Formulation}
To move between complex-valued quantum states and real-valued problem variables, we follow \cite{leung2017gpugrape} and utilize an isomorphic representation for complex vectors and matrices.  We use a \textit{tilde} or the notation $\iso(\cdot)$ to represent isomorphic representations. For a complex-valued vector $\psi \in \mathbf{C}^n$ and matrix $H \in \mathbf{C}^{n \times n}$, we then have, respectively,
\begin{equation}
  \isopsi = \mqty(\Re \psi \\ \Im \psi) \quad \text{and} \quad
  \widetilde{H} = \mqty(\Re H & -\Im H \\ \Im H & \Re H),
\end{equation}
where $\isopsi \in \mathbf{R}^{2n}$ and $\widetilde H \in \mathbf{R}^{2n \times 2n}$.

Since the dynamics involve an evaluation of $\exp(-iH)$ --- i.e. exponentiation of the \textit{generator} --- we also define: 
\begin{equation}
  G(H) := \iso(-iH) = \mqty(\Im H & \Re H \\ -\Re H & \Im H). 
\end{equation}
\noindent
And, since $H$ is linear in $\vb{a}$ and $G$ is a linear function of $H$, we define:
\begin{equation}
  G(\vb{a}) := G(H(\vb{a})) = G(H_0) + \sum_j a_i G(H_i).
\end{equation}


\subsection{Pad\'e Integrators}
The isomorphic dynamics are still in the form of Eqn.~(\ref{eq: quantum-dynamics-naive}):   
\begin{equation}
  f(\isoU_{k+1}, \isoU_k, \vb{a}_k, \Delta t) = \isoU_{k+1} - \exp(G(\vb{a}_k) \Delta t) \ \isoU_k.
\end{equation}
\noindent
The matrix exponential in its present form is a costly operation that does not account for how matrix exponentials are numerically computed in practice. One approach to address this is to use the Pad\'e approximant for the exponential \cite{Moler1978NineteenDW}, which approximates the matrix exponential as 
\begin{equation}
  \exp(A) \approx B^{-1}(A)F(A),
\end{equation}
\noindent
where $B(A)$ and $F(A)$ are truncated power series in $A$ whose the coefficients can be chosen to match $\exp(A)$ up to some desired order.  

We take advantage of this structure in PICO by recognizing that the leading matrix inverse is computationally expensive and not necessary to compute directly since we are enforcing the dynamics constraints implicitly. Thus, we can rewrite the dynamics using the Pad\'e approximant as:
\begin{equation}
\begin{split}
    \vb{P}(\isoU_{k+1}, \isoU_k, \vb{a}_k, \Delta t) := B(\vb{a}_k, \Delta t) \isoU_{k+1} - F(\vb{a}_k, \Delta t) \isoU_k.
\end{split}
\end{equation}

Since we solve most of our problems in a rotating frame without very fast dynamics, we find that the fourth order diagonal \textit{Pad\'e integrator}, denoted $\vb{P}^{(4)}$ is sufficient. It is defined by
\begin{equation}
    B^{(4)}(\vb{a}, \Delta t) := I - \frac{\Delta t}{2} G(\vb{a})  + \frac{\Delta t^2}{12} G(\vb{a}) ^2
\end{equation}
\noindent
and
\begin{equation}
    F^{(4)}(\vb{a}, \Delta t) := I + \frac{\Delta t}{2} G(\vb{a})  + \frac{\Delta t^2}{12} G(\vb{a}) ^2.
\end{equation}

Intuitively, $B$ evolves the state backward a half step in time and $F$ evolves the state forward a half step. Pad\'e approximants have nice properties w.r.t. matrix Lie groups, namely that they are \textit{structure-preserving} \cite{cardoso2001logpade}. In practice, we find that any difference between rollouts under $\vb{P}^{(4)}$ and the matrix exponential is always of lower order than the infidelity of the optimal solution. For problems with particularly fast dynamics, higher-order Pad\'e integrators can be used at very low additional computational cost. 

\subsection{Smooth Solutions} \label{ssec: smooth}
Smooth control pulses are often desirable. To achieve this, we augment the states of the system with the first and second derivatives of the drive parameter so that the knot points are now $z_k = (\isoU_k, \vb{a}_k, \dot{\vb{a}}_k, \ddot{\vb{a}}_k)^\top$ with $\ddot{\vb{a}}$ the new control variable. This procedure is equivalent to using a piecewise cubic spline interpolation of the controls, which is convenient for interpolating trajectories. The new dynamics for this augmented problem are given by
\begin{equation} \label{eq: smooth dynamics}
  f(z_k, z_{k+1}) = \mqty(\vb{P}^{(n)}(\isoU_{k+1}, \isoU_k, \vb{a}_k, \Delta t) \\ 
  \vb{a}_{k+1} - \vb{a}_k - \dot{\vb{a}}_k \cdot \Delta t \\ 
  \dot{\vb{a}}_{k+1} - \dot{\vb{a}}_k - \ddot{\vb{a}}_k \cdot \Delta t).
\end{equation}
\noindent
By adding quadratic regularization costs on $\vb{a}_k$, $\dot{\vb{a}}_k$, and $\ddot{\vb{a}}_k$, and possibly bounding constraints on the velocities $\dot{\vb{a}}_k$ or accelerations $\ddot{\vb{a}}_k$, we can easily control the smoothness of the pulse.

\subsection{Time-Optimal Solutions}

In our framework, it is possible, and very useful, to treat the timestep $\Delta t$ as a decision variable: the knot points are augmented as, e.g., $z_k = (\isoU_k, \vb{a}_k, \Delta t_k)^\top$. It is often necessary to add bound constraints, $\Delta t_{\min} < \Delta t_k < \Delta t_{\max}$, to prevent the solver from taking advantage of discretization errors. In practice, it also helps to constrain all the $\Delta t_k$s to be equal.  

This augmentation adds extra freedom to the optimization problem, allowing the optimal duration of the pulse, which is often not known \textit{a priori}, to be found by the solver. This is referred to as a \textit{free-time} problem. Moreover, we can now add a cost term to the objective of the form $J_{\text{mintime}}(\Delta t_{1:N-1}) = \sum_k \Delta t_k$ and an inequality constraint $\mathcal{F}(U_N) \geq \bar{\mathcal{F}}$ on the final state fidelity which is rigorously enforceable in our method.  This allows us to achieve \textit{minimum-time} solutions for a chosen fidelity, which are helpful for realizing higher-fidelity quantum computations in the presence of decoherence.

\section{Simulation Examples} \label{sec: example}

In this section, we describe the results of applying PICO to a set of three examples of increasing difficulty: a minimum-time single-qubit problem, a two-qubit problem, and a three-qubit problem. In all examples, we use a random initial guess for the controls and the geodesic on $SU(n)$ from the identity to the desired gate as a (dynamically infeasible) initial state trajectory. Except for the single-qubit example, all examples in this paper (including the hardware result) use the smooth solution problem formulation described in Sec.~\ref{ssec: smooth}. Code for all of these examples can be found in the QuantumCollocation.jl GitHub repository \cite{Trowbridge_QuantumCollocation_jl_2023}.

\subsection{Single-Qubit Y-Gate Minimum Time Problem}

As a first example, we consider a single-qubit system defined in QCTRL's user guide \cite{qctrl-time-optimal} (note that we ignore the noise factor and compare against the system ignoring robustness) with the Hamiltonian:
\begin{equation}
    H(\alpha(t), \gamma(t)) = \frac{1}{2} \qty(\gamma(t)\sigma_- + \gamma^*(t)\sigma_+) + \frac{\alpha(t)}{2}\sigma_z.
\end{equation}
\noindent
Where $\alpha \in \mathbf{R}$, $\gamma \in \mathbf{C}$, $\sigma_+$ and $\sigma_-$ are the qubit ladder operators and $\sigma_z$ is the Pauli $z$ matrix.  The goal is to find a time-optimal pulse that enacts a $Y$-gate (i.e. $U_{\text{goal}}=\sigma_y$) while enforcing $|\alpha|< \alpha_\text{max}=2\pi \times 0.1$ MHz and $|\gamma|< \gamma_\text{max}= 2\pi \times 0.3$ MHz.

To achieve this goal we first solve the following \textit{free-time} problem:
\begin{equation} \label{prob: 1qubit}
\begin{split}
    \minimize{z_{1:N}} \quad & J_0(z_{1:N}) = Q \cdot \ell(\isoU_N) + R(\alpha_{1:N-1}, \gamma_{1:N-1}) \\
    \st \quad & \vb{P}^{(4)}\qty(\isoU_{k+1}, \isoU_k, (\alpha_k, \gamma_k), \Delta t_k) = 0 ,\\
    & \isoU_1 = I_{2n} ,\\
    & |\alpha_k|< \alpha_\text{max} ,\\
    & |\gamma_k|^2< \gamma_\text{max}^2 ,\\
    & \Delta t_{\min} < \Delta t_k < \Delta t_{\max} ,
\end{split}
\end{equation}
\noindent
where $z_k = (\isoU_k, \alpha_k, \gamma_k, \Delta t_k)$, $N=100$, $Q = 200$ is a user-specified cost shaping parameter, and $R(\alpha_{1:N-1}, \gamma_{1:N-1})$ is a quadratic regularization objective on the magnitude of the controls (see code for details). 

\begin{figure}[h!]
    \centering
    \includegraphics[width = \linewidth]{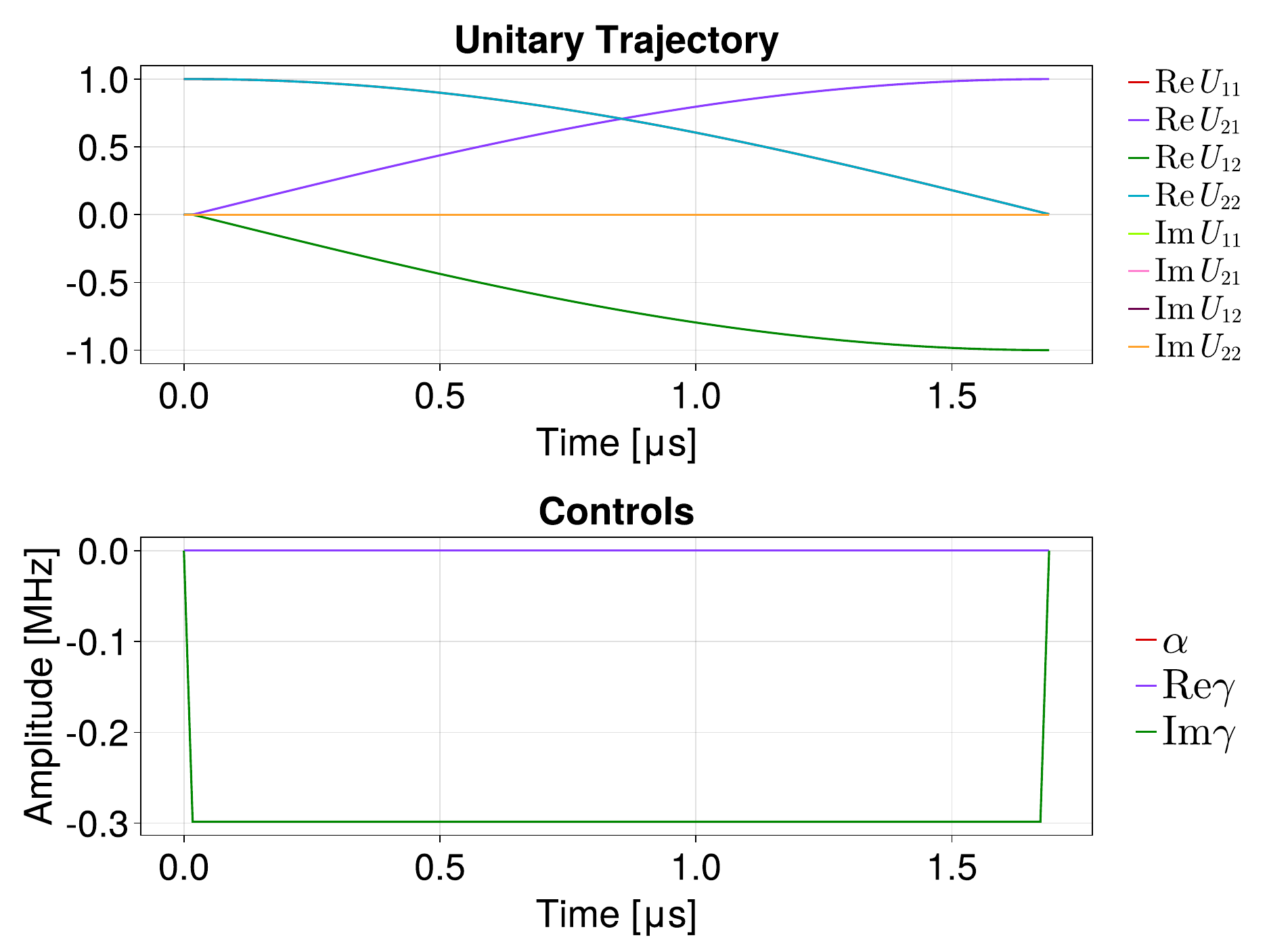}
    \caption{
        \textbf{Single qubit Y-gate minimum time solution.}
        The top plot shows the evolution of the real and imaginary components of the unitary operator over the duration of the pulse.  The bottom plot shows the pulse returned by PICO, where it is observed to have found the analytic \textit{bang-bang} solution on its own ($\alpha$ and $\Re \gamma$ were effectively zeroed out). The optimized pulse has a duration of $1.67 \, \mu \text{s}$ and a final rollout infidelity of $4.72\times 10^{-6}$. We solved the same minimum time problem using Q-CTRL \cite{qctrl-time-optimal} which achieved a duration of $1.68 \, \mu \text{s}$ and infidelity of $1.44 \times 10^{-5}$ as defined in Eqn.~\ref{eq: trace_fid}. Notably, unlike the PICO solution, this solution did not zero out $\alpha$ and $\Re \gamma$.
    }
    \label{fig:single_qubit}
\end{figure}

With an initial solution to \eqref{prob: 1qubit} found by PICO we then warm-start a second problem with a new objective $J = J_0 + D\sum_k \Delta t_k$, and an inequality constraint on the final fidelty, $\mathcal{F}(U_N) \leq \bar{\mathcal{F}}$, to prevent the fidelity from decreasing while we minimize the duration of the pulse. We used $D = 10^{9}$ and $1 - \bar{\mathcal{F}} = 5 \times 10^{-6}$. The final solution has an exponential rollout infidelity of $4.72\times 10^{-6}$ and a duration of $1.67 \ \mu$s.  As can be seen in Fig.~\ref{fig:single_qubit}, PICO has found the analytical \textit{bang-bang} solution. Q-CTRL's proprietary method also finds a comparable solution, which is to be expected for this simple problem, but with PICO we are able to fix the fidelity and solve for the duration.

\begin{figure}[h!]
\centering
\includegraphics[width=\columnwidth]{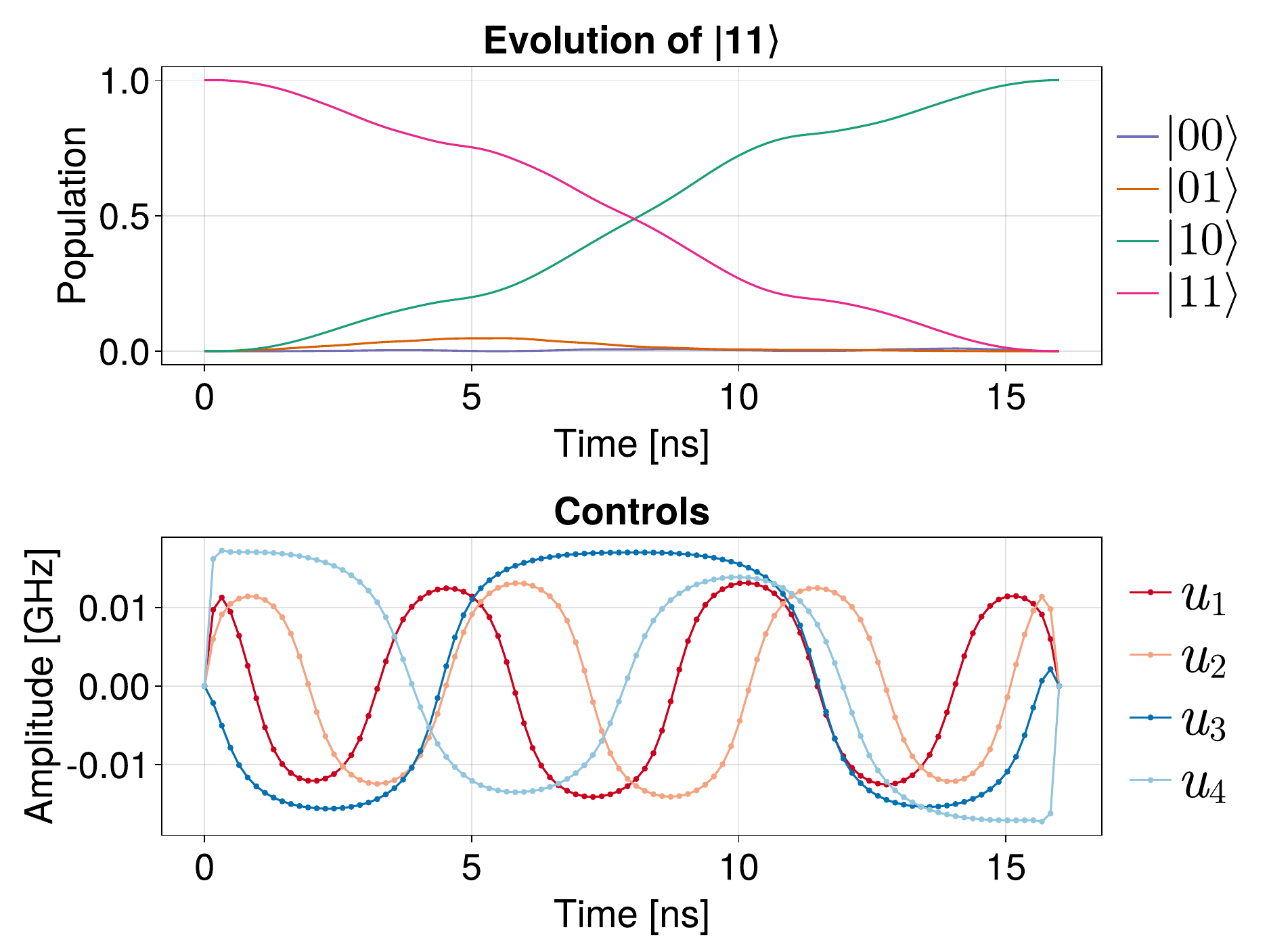}
\caption{
\textbf{Two-qubit CNOT gate.}
Optimized pulse of duration $16$ ns and rollout infidelity of $2.25 \times 10^{-8}$. The top plot shows the evolution of the $\ket{11}$ state under the control pulse, while the bottom plot shows the pulse itself. The dots in the bottom plot correspond to the $N = 100$ knot points of the trajectory.}
\label{twoqubit}
\end{figure}

\subsection{Two-Qubit CNOT Gate Problem}
As a second example, we consider a two-qubit Hamiltonian in the rotating frame with direct qubit drives:
\begin{equation}
    \begin{split}
    H(u(t))  &= g \left(\hat{a}^\dagger \hat{a} \right) \left(\hat{b}^\dagger \hat{b} \right) \\
    & \quad + u_1(t)\left(\hat{a} + \hat{a}^\dagger\right) + i u_2(t)\left(\hat{a} - \hat{a}^\dagger \right) \\ & \quad + u_3(t)\left(\hat{b} + \hat{b}^\dagger\right) + i u_4(t)\left(\hat{b} - \hat{b}^\dagger \right),
\end{split}
\end{equation} 
where $g/{2\pi} = 100$ MHz and $\hat{a}$ and $\hat{b}$ are the annihilation operators of the first and second qubits, respectively. We optimize a CNOT gate using the free-time problem framework subject to the control bounds $|u_{i = 1:4}(t)|/{2\pi} < 20$ MHz. The theoretical minimum CNOT gate time is given by $2\pi/g = 10$ ns, although this may not be achievable under our control bounds. Hence, as a starting point, we solve an initial free time problem with $N = 100$ knot points and an initial guess of $\Delta t_{k = 1:N-1} = 0.1$ ns between knot points, subject to the constraint that all the $\Delta t_k$ between knot points are equal and $0.09 < \Delta t_k < 0.17$ ns. This problem converges to a $15$ ns solution with rollout infidelity $3.67 \times 10^{-8}$ in just $64$ solver iterations. 
Given this solution, we then solve another free-time problem with an initial guess of $\Delta t_k = 0.15$ and bounds $0.135 < \Delta t_k < 0.210$ ns. This problem converges to a $16$ ns solution in around $500$ solver iterations with a rollout infidelity of $2.25 \times 10^{-8}$ as shown in Fig. \ref{twoqubit}. 

This example highlights two of the compelling convergence properties of PICO: 1) dynamically infeasible initial guesses where the state trajectory does match the controls and intermediate dynamics constraint violations have the potential to enable fast convergence and 2) the favorable tail-convergence properties afforded by direct methods result in many $9$s of fidelity in far fewer iterations than indirect methods.



\subsection{Three-Qubit SWAP Gate Problem} 

In this example, we consider the three-qubit Hamiltonian
\begin{equation}
    \begin{split}
        H(u(t)) = 2\pi H_0 + \sum_{j=1}^3 u_j(t) \hat{a}_j + u_j(t)^* \hat{a}_j^\dagger,
    \end{split}
\end{equation}

\noindent
where $\mathbf{u}(t) \in \mathbf{C}^3$, $\hat a_j$ is the annihilation operator on the $j$-th qubit, and the drift term is,
\begin{equation}
\begin{split}
    H_0 = &\sum_{j=1}^3 (\omega_j -\omega_d) \hat{a}_j^\dagger \hat{a}_j - \frac{\xi}{2} \hat{a}_j^\dagger \hat{a}_j^\dagger \hat{a}_j \hat{a}_j \\
    &+ J_{12} \qty(\hat{a}_1^\dagger \hat{a}_2 + \hat{a}_1 \hat{a}_2^\dagger) + J_{23} \qty(\hat{a}_2^\dagger \hat{a}_3 + \hat{a}_2 \hat{a}_3^\dagger) ,
\end{split}
\end{equation}
\noindent
where (in units of GHz) $\omega_1 = 5.18$, $\omega_2 = 5.12$, $\omega_3 = 5.06 $, $\omega_d = 5.12$, $\xi = 0.34$, and $J_{12} = J_{23} = 5.0 \times 10^{-3}$. We used 500 timesteps, where $T = 200 \, \text{ns}$, $\Delta t = T / 500$, $\Delta t_{\min} = 0.5 \Delta t$, and $\Delta t_{\max} = \Delta t$. We also enforced box constraints of $0.04$ GHz on $\abs{\Re u^j} / 2\pi$ and $\abs{\Im u^j} / 2\pi$. The goal is to enact a $1-3$ SWAP gate, i.e.
\begin{equation}
\begin{split}
    U_{\text{goal}} &=
        \op{0}{0} \otimes I \otimes \op{0}{0}\\
        &+ \op{1}{0} \otimes I \otimes \op{0}{1}\\
        &+ \op{0}{1} \otimes I \otimes \op{1}{0}\\
        &+ \op{1}{1} \otimes I \otimes \op{1}{1}.
\end{split}
\end{equation}

\begin{figure}[h!]
    \centering
    \includegraphics[width = \linewidth]{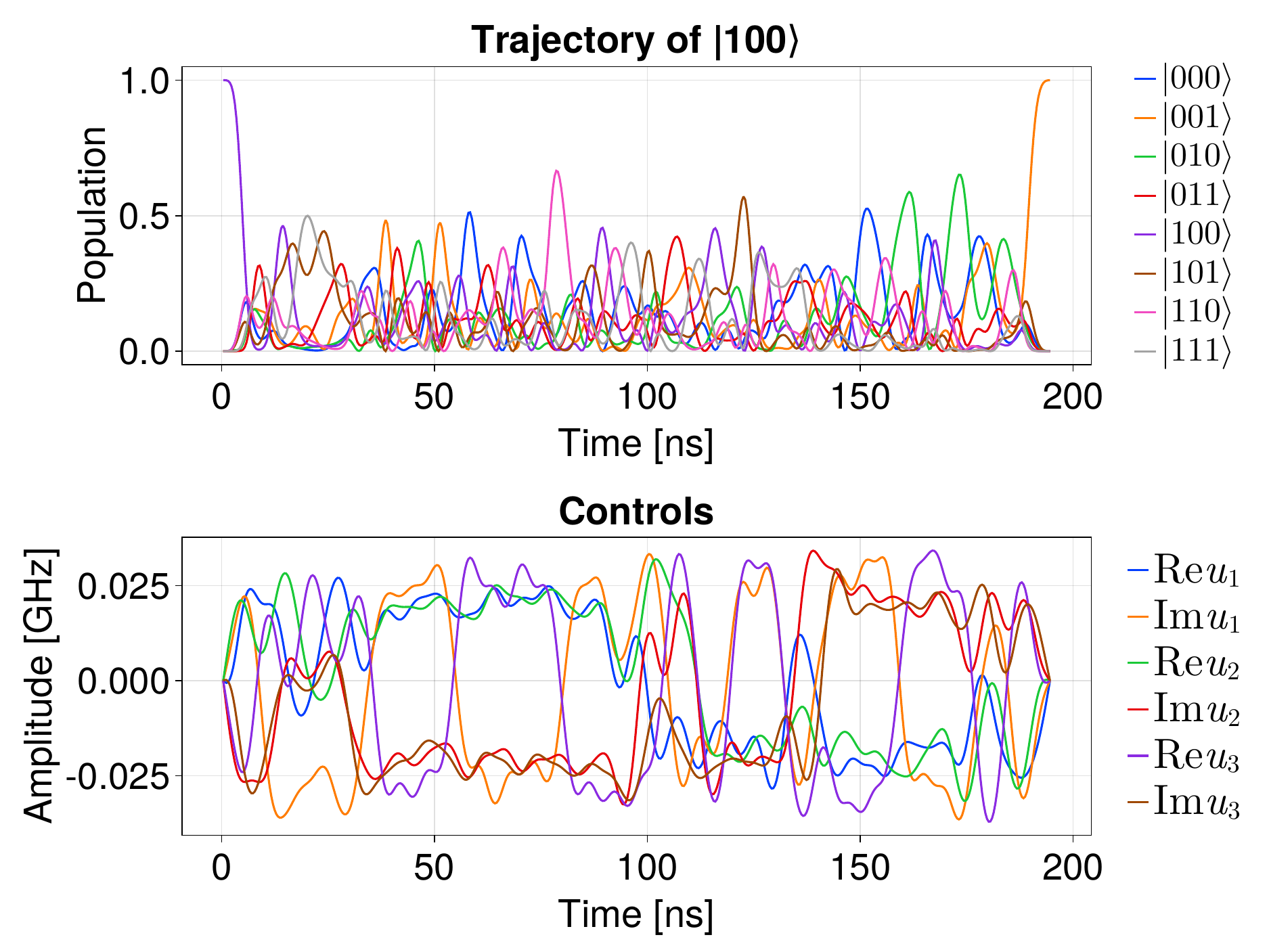}
    \caption{
        \textbf{Three-qubit 1-3 SWAP gate pulse.}
        The top plot shows the populations of a $\ket{100}$ state, evolved via the exponential, with the controls optimized for the target 1-3 SWAP gate. The bottom plot shows the controls.  This pulse achieved solver infidelity of $3.4 \times 10^{-5}$ and exponential rollout infidelity of $8.6\times 10^{-5}$ with a duration of $194.56 \, \text{ns}$.
    }
    \label{fig:three_qubit}
\end{figure}

This problem was solved using the smooth solution dynamics in \eqref{eq: smooth dynamics} and knot points $z_k = (\isoU_k, \mathbf{u}_k, \dot{\mathbf{u}}_k, \ddot{\mathbf{u}}_k, \Delta t_k)^\top$.  The real and imaginary components of $\ddot u_k$ were constrained to have absolute values less than $0.05$ to enforce smoothness.  The solution is shown in Fig. \ref{fig:three_qubit}; code for this solution can be found in the QuantumCollocation.jl GitHub repository.

\section{Hardware Results} \label{sec: hwr}

In applying PICO to a hardware system a few other details were needed.  Specifically, we were interested in a state transfer instead of a unitary gate. 
Additionally, in order to simulate the bosonic oscillator component of our hardware system, we truncated its infinite-dimensional Hilbert space. To ensure PICO did not take advantage of the artificial nonlinearity created by this truncation, we prevented the population of the highest levels in our truncated model by imposing high $L_1$ regularization costs on them in the problem objective. The result is a success in both simulation, as can be seen in Fig. \ref{hardware_figure_p1}, and in the outcome of the hardware experiment, as described below. 


\subsection{Experimental Outcome}

\begin{figure}[t]
\includegraphics[width=0.4832\textwidth]{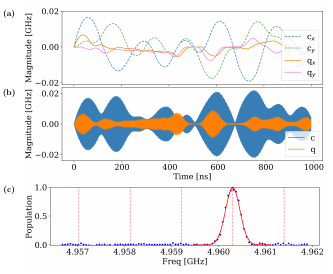}
\caption{ \textbf{PICO pulse experimental results for a target state of $\ket{g1}$}. \textbf{(a)} Initial controls $u_j(t)$ are solved for in a rotating frame at the qubit and cavity frequencies, with a Hamiltonian dominated by the dispersive shift $\chi$ interaction term. Each of the cavity (c) and qubit (q) pulses are split into their real (x) and imaginary (y) quadratures to allow for effectively complex controls while solving with only real numbers. \textbf{(b)} The quadratures of the cavity and qubit control pulses are recombined in the lab frame before being sent to the experiment's control instrumentation. This also involves modulating at the cavity or qubit frequency. The sampling rate of the DAC used to generate the pulses is 16 GSa/s, with 8 bits of amplitude resolution.
\textbf{(c)} Resolved qubit spectroscopy of the state produced by the control pulses. Dashed lines indicate locations of photon-number resolved peaks, corresponding to $0, 1, \ldots, 4$ photons from right to left. The final system state matches the target $\ket{g1}$ with fidelity $0.988 \pm 0.011$.
} 
\label{hardware_figure_p2}
\end{figure}

Control pulses were tested on a 3D superconducting circuit-cavity QED platform that features a nonlinear transmon qubit coupled to a bosonic oscillator cavity and a readout cavity~\cite{chakramhe2022blockade}. 
This system implements the following Hamiltonian,
\begin{equation}
\begin{split}
    H(u_q(t), u_c(t)) = &\hspace{0.15cm} \omega_{q} \hat{b}^\dag \hat{b} + \frac{\alpha}{2} \hat{b}^\dag \hat{b} (\hat{b}^\dag \hat{b} - 1) + \chi \hat{a}^\dag \hat{a} \hat{b}^\dag \hat{b} \\
     & + \omega_c \hat{a}^\dag \hat{a}  + \frac{k}{2} \hat{a}^\dag \hat{a} (\hat{a}^\dag \hat{a} - 1) \\
     & + u_q(t)(\hat{b} + \hat{b}^\dag) + u_c(t)(\hat{a} + \hat{a}^\dag),
\end{split}
\end{equation}
where $\hat{b}$ is the annihilation operator for the transmon, $\omega_q = 2\pi \times 4.96$ GHz is the qubit $\ket{g}-\ket{e}$ transition frequency, $\alpha = 2\pi \times -0.143$ GHz is the anharmonicity, $\hat{a}$ is the annihilation operator of the cavity, $\omega_c = 2\pi \times 6.22$ GHz is the cavity frequency, $\chi = 2\pi \times -1.13$ MHz is the dispersive interaction term, and $k = 2\pi \times 4.2$ kHz is the self-Kerr of the cavity. Controls $u_q(t)$ and $u_c(t)$ are time-dependent input pulses that determine the dynamics of the system. Pulses were generated by modeling three levels of the transmon and 14 levels of the cavity with $L_1$ costs on the highest levels.

To measure the performance of a set of control pulses, we perform photon-number-resolved qubit spectroscopy. Due to the $\chi$ dispersive shift term, individual-cavity photon-number populations change the qubit frequency and form distinguishable peaks~\cite{schuster2007resolving}. As an example, we test solutions for state preparation of $\ket{g1}$ (transmon in the ground state, one photon in the cavity), and obtain an experimental fidelity of $0.988 \pm 0.011$, as shown in Fig.~\ref{hardware_figure_p2}, with the error obtained from simulating the experiment process. This fidelity is in line with state-of-the-art results~\cite{ecd2022, sivak2022highfidfock}, and is close to the simulated fidelity under decoherence of $0.997.$ We attribute the discrepancy to slight inaccuracies or fluctuations in experimental system parameters, as well as quantization errors in the control electronics. Additional simulations showing the expected evolution and state populations at different time slices for the duration of the pulse are shown in Fig.~\ref{hardware_figure_p1} in the form of reduced traces on the transmon and cavity and Wigner tomography on the cavity.

\section{Conclusions and Future Work}
We have introduced PICO, a direct collocation method for quantum optimal control. By treating both the states and controls as decision variables, in contrast to indirect methods, PICO is able to achieve exceptional performance by leveraging state-of-the-art sparse nonlinear programming solvers like IPOPT.  As a direct method, which can handle general nonlinear constraints on both the states and controls, PICO demonstrably outperforms existing indirect methods. In particular, it can solve minimum-time problems with constraints on the final state fidelity, allowing it to minimize pulse durations without sacrificing performance.   PICO's other capabilities --- e.g. a free-time problem formulation that allows the solver to find the optimal pulse duration --- and experimental results, both in simulation and on hardware, show that is a powerful approach.  PICO is open-source and available via a registered Julia package: QuantumCollocation.jl.

There are a number of exciting directions for future work. One avenue is to develop a custom nonlinear solver that takes advantage of the $SU(n)$ structure inherent in our dynamics: solvers, such as IPOPT, use techniques \cite{nocedal1999numerical, boyd2004convex, wachter2006ipopt} that could be specialized for this setting. Another direction, which we view as crucial to hardware applications, is utilizing \textit{iterative learning control} \cite{bristow2006ILCsurvey} combined with PICO to correct model-mismatch errors. 

\newpage

\section*{Acknowledgements}
For helpful discussions and for providing the three-qubit Hamiltonian we thank Anders Petersson and Stefanie Guenther at Lawrence Livermore National Laboratory. We used Makie.jl and matplotlib to generate our figures.  We would also like to thank the entire Julia software community which made these results possible.

\bibliography{bibliography}

\begin{thebibliography}{10}
\providecommand{\url}[1]{#1}
\csname url@samestyle\endcsname
\providecommand{\newblock}{\relax}
\providecommand{\bibinfo}[2]{#2}
\providecommand{\BIBentrySTDinterwordspacing}{\spaceskip=0pt\relax}
\providecommand{\BIBentryALTinterwordstretchfactor}{4}
\providecommand{\BIBentryALTinterwordspacing}{\spaceskip=\fontdimen2\font plus
\BIBentryALTinterwordstretchfactor\fontdimen3\font minus
  \fontdimen4\font\relax}
\providecommand{\BIBforeignlanguage}[2]{{%
\expandafter\ifx\csname l@#1\endcsname\relax
\typeout{** WARNING: IEEEtran.bst: No hyphenation pattern has been}%
\typeout{** loaded for the language `#1'. Using the pattern for}%
\typeout{** the default language instead.}%
\else
\language=\csname l@#1\endcsname
\fi
#2}}
\providecommand{\BIBdecl}{\relax}
\BIBdecl

\bibitem{manchesterTrajOpt}
Z.~Manchester and S.~Kuindersma, ``Robust direct trajectory optimization using
  approximate invariant funnels,'' \emph{Autonomous Robots}, vol.~43, no.~2,
  pp. 375--387, 2018.

\bibitem{altro}
T.~A. Howell, B.~E. Jackson, and Z.~Manchester, ``Altro: A fast solver for
  constrained trajectory optimization,'' in \emph{2019 IEEE/RSJ International
  Conference on Intelligent Robots and Systems (IROS)}, 2019, pp. 7674--7679.

\bibitem{howell2023direct}
T.~A. Howell, C.~Fu, and Z.~Manchester, ``Direct policy optimization using
  deterministic sampling and collocation,'' 2023.

\bibitem{jackson2021attitude}
B.~E. Jackson, K.~Tracy, and Z.~Manchester, ``Planning with attitude,''
  \emph{IEEE Robotics and Automation Letters}, vol.~6, no.~3, pp. 5658--5664,
  2021.

\bibitem{underactuated}
\BIBentryALTinterwordspacing
R.~Tedrake, \emph{Underactuated Robotics}, 2023. [Online]. Available:
  \url{https://underactuated.csail.mit.edu}
\BIBentrySTDinterwordspacing

\bibitem{khaneja2005grape}
\BIBentryALTinterwordspacing
N.~Khaneja, T.~Reiss, C.~Kehlet, T.~Schulte-Herbrüggen, and S.~J. Glaser,
  ``Optimal control of coupled spin dynamics: design of nmr pulse sequences by
  gradient ascent algorithms,'' \emph{Journal of Magnetic Resonance}, vol. 172,
  no.~2, pp. 296--305, 2005. [Online]. Available:
  \url{https://www.sciencedirect.com/science/article/pii/S1090780704003696}
\BIBentrySTDinterwordspacing

\bibitem{caneva2011crab}
\BIBentryALTinterwordspacing
T.~Caneva, T.~Calarco, and S.~Montangero, ``Chopped random-basis quantum
  optimization,'' \emph{Physical Review A}, vol.~84, no.~2, aug 2011. [Online].
  Available: \url{https://doi.org/10.1103\%2Fphysreva.84.022326}
\BIBentrySTDinterwordspacing

\bibitem{hargraves1987dircol}
C.~Hargraves and S.~Paris, ``Direct trajectory optimization using nonlinear
  programming and collocation,'' \emph{AIAA J. Guidance}, vol.~10, pp.
  338--342, 07 1987.

\bibitem{wachter2006ipopt}
A.~W{\"a}chter and L.~T. Biegler, ``On the implementation of an interior-point
  filter line-search algorithm for large-scale nonlinear programming,''
  \emph{Mathematical programming}, vol. 106, pp. 25--57, 2006.

\bibitem{Trowbridge_QuantumCollocation_jl_2023}
\BIBentryALTinterwordspacing
A.~Trowbridge and A.~Bhardwaj, ``{QuantumCollocation.jl},'' Feb. 2023.
  [Online]. Available:
  \url{https://github.com/aarontrowbridge/QuantumCollocation.jl}
\BIBentrySTDinterwordspacing

\bibitem{Hargraves87a}
C.~R. Hargraves and S.~W. Paris, ``Direct {{Trajectory Optimization Using
  Nonlinear Programming}} and {{Collocation}},'' \emph{J. Guidance}, vol.~10,
  no.~4, pp. 338--342, 1987.

\bibitem{Betts92}
\BIBentryALTinterwordspacing
J.~T. Betts and W.~P. Huffman, ``Application of sparse nonlinear programming to
  trajectory optimization,'' \emph{Journal of Guidance, Control, and Dynamics},
  vol.~15, no.~1, pp. 198--206, Jan. 1992. [Online]. Available:
  \url{https://arc.aiaa.org/doi/10.2514/3.20819}
\BIBentrySTDinterwordspacing

\bibitem{Pardo15a}
\BIBentryALTinterwordspacing
D.~Pardo, L.~M{\"o}ller, M.~Neunert, A.~W. Winkler, and J.~Buchli, ``Evaluating
  direct transcription and nonlinear optimization methods for robot motion
  planning,'' pp. 1--9, Apr. 2015. [Online]. Available:
  \url{http://arxiv.org/pdf/1504.05803v1.pdf}
\BIBentrySTDinterwordspacing

\bibitem{Posa16a}
M.~Posa, S.~Kuindersma, and R.~Tedrake, ``Optimization and stabilization of
  trajectories for constrained dynamical systems,'' in \emph{Proceedings of the
  {{International Conference}} on {{Robotics}} and {{Automation}}
  ({{ICRA}})}.\hskip 1em plus 0.5em minus 0.4em\relax {Stockholm, Sweden}:
  {IEEE}, 2016, pp. 1366--1373.

\bibitem{Betts98a}
\BIBentryALTinterwordspacing
J.~T. Betts, ``Survey of {{Numerical Methods}} for {{Trajectory
  Optimization}},'' \emph{Journal of Guidance, Control, and Dynamics}, vol.~21,
  no.~2, pp. 193--207, Mar. 1998. [Online]. Available:
  \url{http://arc.aiaa.org/doi/10.2514/2.4231}
\BIBentrySTDinterwordspacing

\bibitem{leung2017gpugrape}
\BIBentryALTinterwordspacing
N.~Leung, M.~Abdelhafez, J.~Koch, and D.~Schuster, ``Speedup for quantum
  optimal control from automatic differentiation based on graphics processing
  units,'' \emph{Phys. Rev. A}, vol.~95, p. 042318, Apr 2017. [Online].
  Available: \url{https://link.aps.org/doi/10.1103/PhysRevA.95.042318}
\BIBentrySTDinterwordspacing

\bibitem{boulder_opal1}
\BIBentryALTinterwordspacing
H.~Ball, M.~J. Biercuk, A.~R.~R. Carvalho, J.~Chen, M.~Hush, L.~A.~D. Castro,
  L.~Li, P.~J. Liebermann, H.~J. Slatyer, C.~Edmunds, V.~Frey, C.~Hempel, and
  A.~Milne, ``Software tools for quantum control: improving quantum computer
  performance through noise and error suppression,'' \emph{Quantum Science and
  Technology}, vol.~6, no.~4, p. 044011, 2021. [Online]. Available:
  \url{https://doi.org/10.1088/2058-9565/abdca6}
\BIBentrySTDinterwordspacing

\bibitem{boulder_opal2}
Q-CTRL, ``Boulder {O}pal,'' https://q-ctrl.com/boulder-opal, 2023, [Online].

\bibitem{Moler1978NineteenDW}
C.~B. Moler and C.~V. Loan, ``Nineteen dubious ways to compute the exponential
  of a matrix, twenty-five years later,'' \emph{SIAM Rev.}, vol.~45, pp. 3--49,
  1978.

\bibitem{cardoso2001logpade}
\BIBentryALTinterwordspacing
J.~Cardoso and F.~{Silva Leite}, ``Theoretical and numerical considerations
  about padé approximants for the matrix logarithm,'' \emph{Linear Algebra and
  its Applications}, vol. 330, no.~1, pp. 31--42, 2001. [Online]. Available:
  \url{https://www.sciencedirect.com/science/article/pii/S0024379501002518}
\BIBentrySTDinterwordspacing

\bibitem{qctrl-time-optimal}
Q-CTRL, ``Boulder {O}pal documentation: {H}ow to find time-optimal controls,''
  /boulder-opal/application-notes/designing-robust-configurable-parallel-gates-for-large-trapped-ion-arrays,
  2022, [Online; accessed 29-April-2023].

\bibitem{chakramhe2022blockade}
S.~Chakram, K.~He, A.~Dixit, A.~Oriani, R.~Naik, N.~Leung, H.~Kwon, W.~Ma,
  L.~Jiang, and D.~Schuster, ``Multimode photon blockade,'' \emph{Nature
  Physics}, vol.~18, no.~1, pp. 879--884, 2022.

\bibitem{schuster2007resolving}
D.~Schuster, A.~Houck, J.~Schreier, A.~Wallraff, J.~Gambetta, A.~Blais,
  L.~Frunzio, J.~Majer, B.~Johnson, M.~Devoret \emph{et~al.}, ``Resolving
  photon number states in a superconducting circuit,'' \emph{Nature}, vol. 445,
  no. 7127, pp. 515--518, 2007.

\bibitem{ecd2022}
A.~Eickbush, V.~Sivak, A.~Ding, S.~Elder, S.~Jha, J.~Venkatraman, B.~Royer,
  S.~Girvin, R.~Schoelkopf, and M.~Devoret, ``Fast universal control of an
  oscillator with weak dispersive coupling to a qubit,'' \emph{Nature Physics},
  vol.~18, no.~1, pp. 1464--1469, 2022.

\bibitem{sivak2022highfidfock}
V.~Sivak, A.~Eickbusch, H.~Liu, B.~Royer, I.~Tsioutsios, and M.~Devoret,
  ``Model-free quantum control with reinforcement learning,'' \emph{Physical
  Review X}, vol.~12, no.~1, p. 011059, 2022.

\bibitem{nocedal1999numerical}
J.~Nocedal and S.~J. Wright, \emph{Numerical optimization}.\hskip 1em plus
  0.5em minus 0.4em\relax Springer, 1999.

\bibitem{boyd2004convex}
S.~Boyd, S.~P. Boyd, and L.~Vandenberghe, \emph{Convex optimization}.\hskip 1em
  plus 0.5em minus 0.4em\relax Cambridge university press, 2004.

\bibitem{bristow2006ILCsurvey}
D.~A. Bristow, M.~Tharayil, and A.~G. Alleyne, ``A survey of iterative learning
  control,'' \emph{IEEE control systems magazine}, vol.~26, no.~3, pp. 96--114,
  2006.

\end{thebibliography}

\end{document}